\documentclass[pre,twocolumn,showpacs]{revtex4}

\usepackage{graphicx}

\begin{document}

\title{Experimental Realization of the Fuse Model }
\author{D. R. Otomar} 
 \author{I.~L. Menezes-Sobrinho} 
 \email{ilima@ufv.br}
\author{M. S. Couto}
\affiliation{Departamento de F\'{\i}sica - Universidade Federal de Vi\c{c}osa, 36571-000,
Vi\c{c}osa,
Minas Gerais, Brazil}

\begin{abstract}
In this work, we present an experimental investigation of the
fuse model. Our main goal was to study the influence of the disorder on the fracture process. The experimental apparatus used consisted of an $L\times L$ square lattice with fuses placed on each bond of the lattice. Two types of materials were used as
fuses: copper and steel wool wires. The lattice composed only by copper wires varied from a weakly disordered system to a strongly disordered one. The lattice formed only by steel wool wires corresponded to a strongly disordered one. The experimental procedure consisted of applying a potential difference V to the lattice and measuring the respective current I. The characteristic function $I(V)$ obtained was investigated in order to find the scaling law dependence of the voltage and the current on the system size $L$ when the disorder was changed. Our results show that the scaling laws are only verified for the disordered regime.

\end{abstract}


\maketitle
When a material is submitted to a load (static or dynamic) cracks can emerge inside it, which may lead to a catastrophic rupture. The fracture of a material can be classified in two categories: brittle and ductile \cite{livro}. A Brittle fracture is characterized by a large crack which leads to the final rupture of the material. In the ductile fracture, small uncorrelated cracks appear randomly, weakening the material and thus causing its failure. Ductile fracture involves large, nonlinear deformations, whereas brittle fracture does not.

A factor that influences enormously the breakdown process is the disorder. It can be greatly amplified during the propagation of the crack and has a strong influence on the roughness of the fracture surface. In any rupture experiment the
presence of disorder is fundamental in the dynamics of crack propagation. It is naturally present
in all materials and comes from a variety of sources. Intense theoretical and
experimental efforts have been done trying to understand the process of crack formation and 
propagation in disordered materials \cite{livro,isma,papel,pre,alex,arc}. 

Many models have been proposed, trying to explain the breakdown process in disordered materials. Among the most studied is the random fuse model (RFM) \cite{arc,arc1,dela,nuk,hansen1,hansen2}. This model consists of a square lattice,  with one fuse at each bond, subjected to an increasing voltage. The fuse model is the electric counterpart of the elastic problem, where force is replaced by current and displacement by voltage.

In the RFM proposed by Arcangelis and Herrmann \cite{arc} all fuses had identical characteristics. Disorder was introduced via dilution, i.e. by having a certain fraction of fuses removed, already before the breaking process starts. In another work, Arcangelis and Herrmann \cite{arc1} studied a fuse lattice with all fuses having the same resistance and a threshold current chosen randomly according to a probability distribution. The current flowing in each fuse was calculated by solving Kirchhoff equations. They performed numerical simulations of the RFM in order to show how the amount of disorder influences the breaking process of a disordered material. They detected two types of regimes: an initial disorder controlled regime in which the current increases monotonically with the applied voltage, and a catastrophic regime in which, for small quenched disorder, the external voltage could also decrease and statistical fluctuations were stronger.
In the first regime the burning of the fuses produces small and isolated cracks which weaken the system and thus causes its rupture. In this regime the current $I$, voltage $V$ and the lattice size $L$ obey the following scaling law 
\begin{equation}
I=L^\alpha(VL^{-\beta}),
\label{eq1}
\end{equation}
where $\alpha\sim0.90$ and $\beta\sim0.84$.
In the catastrophic regime the rupture of the system is due to the formation of a single macroscopic crack which percolates the system. In this regime the data do not seem to obey the scaling function given by Eq. (\ref{eq1}).

The RFM was also investigated in two \cite{eira,ste,hansen1} and three dimensions \cite{rai,bat} to evaluate the roughness exponent of the fracture surface of a disordered material. 
In the RFM the burning of a fuse does not depend on the previous state of the system.  It has no memory and, therefore, the rupture process of the lattice does not have any dynamics but only an irreversible process of rupture without any time scale. 

It is known experimentally that the dynamical aspects are very important in the rupture process of a material. In order to study the dynamical aspects involved in the fracture process of a disordered material the dynamical thermal fuse model \cite{van,sorn} was proposed. In this model an additional temperature field was introduced, in terms of which the rupture criterion was defined. 

The  dynamical thermal fuse model was experimentally investigated by Lamaign\`{e}re {\it et al}. \cite{lamag}. They studied the electrical breakdown of a polymer-particle composite in which an isolating polymer matrix is filled with conducting particles. The particle to particle contacts evolve due to thermal expansion of the matrix as a function of the applied current. It was verified that for a fixed current $I$, above a critical value $I_c$, the electrical resistance $R$ increases as a function of time as a power law.

In this paper we present an experimental realization of the fuse model. The main objective is to study the influence of the disorder in the rupture process of a material.

In our experiment we used a printed circuit board of size 1 m $\times$ 1 m. This
plate was divided in small square regions of 1 cm $\times$ 1 cm across which the
fuses were soldered. Two bus bars were placed across the top and the bottom of
the plate and a voltage difference $V$ was applied between the bars. This
procedure allowed us to make an $L\times L$ square lattice of maximun size
$L=28$, containing a total of 1624 wires. Figure 1 shows a schematic
representation of the lattice.  We used two different types of wires as
fuses: cooper (with a diameter of $0.031$ mm and a conductivity of $5.8~\times
~10^{-7}~ \Omega^{-1} \mbox {m}^{-1}$) and steel wool (with several different
cross sections and conductivity of $5.6\times 10^{-8}~\Omega^{-1}
\mbox{m}^{-1}$). 
\begin{figure}[hbt]
\begin{center}
\resizebox{5cm}{!}{\includegraphics{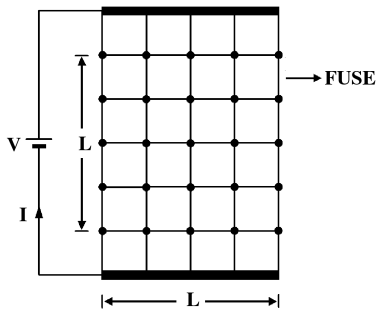}}
\end{center}
\vspace {-0.7cm}
\caption{Schematic representation of the network.}
\label{fig1}
\end{figure}

A lattice formed only by cooper wires of almost equal length and cross section is what we call a weakly disordered lattice (even though the wires are cut with the same length the soldering process introduces very small variations on their length and therefore this lattice is not considered as being completely ordered). A lattice formed only by steel wool wires corresponds to a strongly disordered lattice. Two different manners were used to alter the disorder of the system. For systems composed of copper wires only, disorder was increased by replacing some of the wires by two, three or four wires connected in parallel and by wires of different lengths, placed in bonds chosen at random. Another way of increasing the disorder was to replace some of the copper wires by steel wool wires, again, placed in bonds chosen at random. The percentage of wires replaced defines the disorder $D$ of the lattice. After mounted, the network was submitted to an external voltage, which was varied from zero until the rupture of the lattice in steps of $0.10$ V. For each voltage we registered the value of the total current flowing through the network. The individual current value at each fuse could not be measured.  We consider the following system sizes: $L=7,~ 14,~ 20$ and $28$. In order to verify the reproducibility of the data each experiment was repeated three times.                                                                                                                                                

In order to determine the influence of the disorder on the rupture process, we mounted a lattice of size $28\times28$ and several degrees of disorder. Figure 2 presents the characteristic function $I(V)$ measured for $D$ ranging from $0\%$ to $100\%$ for a lattice composed of copper wires only. 
\begin{figure}[hbt]
\begin{center}
\resizebox{6cm}{!}{\includegraphics{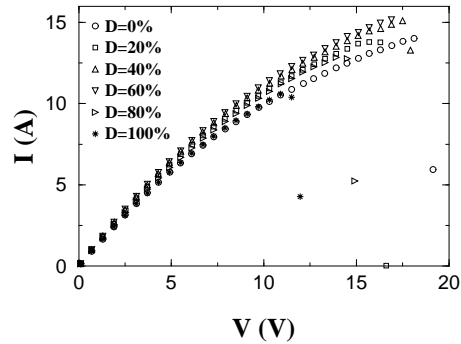}}
\end{center}
\vspace {-0.7cm}
\caption{Characteristic function $I(V)$ measured for several degrees of disorder in systems composed of copper wires only.}
\label{fig2}
\end{figure}

Figure 3 shows the plot of the maximum mean current, $\langle I_{MAX}\rangle$, supported by the network as a function of $D$. Taking into account the error bars, we can distinguish two different regimes in Fig. 3: one for $D<60\%$ and another for $D>60\%$. 
\begin{figure}[hbt]
\begin{center}
\resizebox{6cm}{!}{\includegraphics{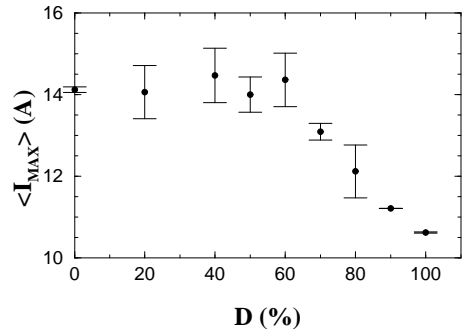}}
\end{center}
\vspace {-0.7cm}
\caption{Maximum mean current $\langle I_{MAX}\rangle$ {\it versus} disorder $D$. Each value was averaged over three samples.}
\label{fig3}
\end{figure}
For $D<60\%$ the current is approximately constant with the increase of the disorder. In this regime the rupture process does not depend on the degree of disorder and the rupture of the network is due to a single crack which rapidly percolates the network. This is a characteristic of brittle fractures. 

For $D>60\%$ the rupture process is controlled by disorder and therefore the rupture of the network proceeds slowly. Initially some bonds (wires) are burned out, provoking the appearance of small and isolated cracks in the system, which causes a decrease of the current that passes through the network. In the next step, new small cracks are created in the network and the fusion of these small cracks produces a large crack which percolates the network. This rupture process is similar to what happens in ductile fracture where the size of the crack that percolates the system is larger than the system size $L$.

The behavior of the function $I(V)$ for different system sizes obtained when using only copper wires is shown in Fig. 4. In this Figure one can see that the electrical current increases as the system size $L$ increases. 
 \begin{figure}[hbt]
\begin{center}
\resizebox{8cm}{!}{\includegraphics{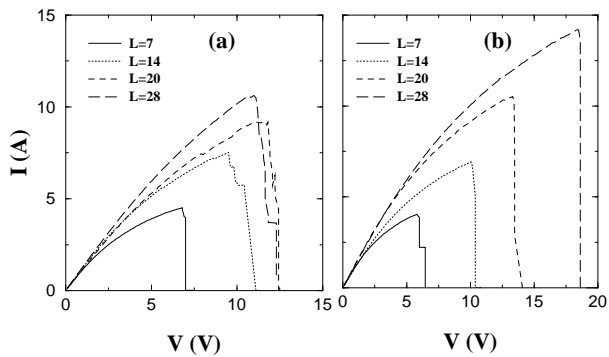}}
\end{center}
\vspace {-0.7cm}
\caption{Current $I$ as a function of the voltage $V$ for different system sizes and two degrees of disorder in systems composed of copper wires only. (a) for  $D=100\%$ and (b) for  $D=0\%$.}
\label{fig4}
\end{figure}
To find the scaling law dependence of the voltage and current on the system size for these two degrees of disorder we attempted to collapse the function $I(V)$ for different system sizes. Our results are presented in Fig. 5. In the strongly disordered regime (Fig. 5(a)) the data obey the scaling of Eq. (\ref {eq1}). The best collapse was found for $\alpha=0.92$ and $\beta=0.87$. In the catastrophic regime (Fig. 5(b)) the data does not obey the scaling of Eq. (\ref {eq1}). Therefore, it was impossible to collapse the data on a single curve. A similar behavior was also obtained by Arcangelis and Herrmann \cite{arc1}. 
\begin{figure}[hbt]
\begin{center}
\resizebox{8cm}{!}{\includegraphics{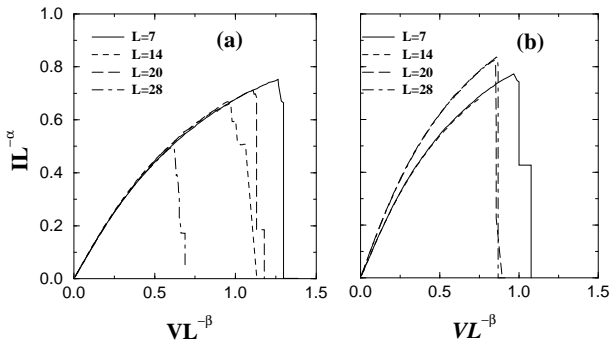}}
\end{center}
\vspace {-0.7cm}
\caption{Plot of the scaling relation $I/L^\alpha$ versus $V/L^\beta$ for copper wires. (a) for $D=100\%$ and (b) for  $D=0\%$. }
\label{fig5}
\end{figure}

Figure 6(a) shows the plot of the function $I(V)$, using steel wool wires, for different system sizes. The scaling law dependence of the voltage and current on the system size is showed in Fig 6(b). In this case the best collapse was found for $\alpha=0.82$ and $\beta=0.87$. The results indicate that the collapse of the $I(V)$ curves does not depend on the type of material, although it depends on the disorder of the network. Our results suggest also that the value of the exponent $\alpha$ depends on the nature of the material employed.

\begin{figure}[hbt]
\begin{center}
\resizebox{8cm}{!}{\includegraphics{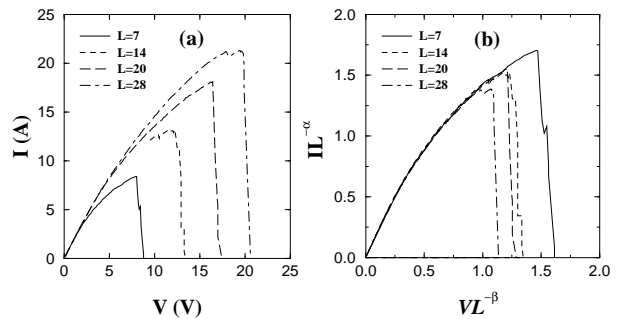}}
\end{center}
\vspace {-0.7cm}
\caption{Results obtained for steel wool wires. (a) plot of the function $I(V)$ and (b) scaling relation $I/L^\alpha$ versus $V/L^\beta$. }
\label{fig6}
\end{figure}

Figure 7 shows the behavior of the total number $\langle n\rangle$ of bonds cut during the whole breaking process as a function of size $L$, for the weakly disordered and the strongly disordered cases. For the strongly disordered case, our results suggest that $\langle n\rangle$ scales with the system size as $L^{1.14}$. The value of the exponent $1.14$, is smaller than the one obtained by Arcangelis and Herrmann, which is 1.71 \cite{arc1}. However, in Refs. 7 and 11 it has been shown that the value of the exponent depends on the degree of disorder of the lattice.  It is possible that the degree of disorder present in our experimental set-up is different from those in Ref. 7. Also, the small lattice size used here may be affecting the value of the exponent. Nevertheless, our experimental observation indicates that in this regime the breaking of the network is due to the fusion of small cracks. Therefore, the final size $\langle n\rangle$ of the crack percolating the system is larger than the system size $L$. For the weakly disordered case, our results suggest the relation $\langle n\rangle \simeq L^{1.02}$, which is approximately linear. This result is in agreement with the picture that in the weakly disordered regime the system is broken by one single macroscopic crack \cite{livro,arc1}.  

\begin{figure}[hbt]
\begin{center}
\resizebox{6cm}{!}{\includegraphics{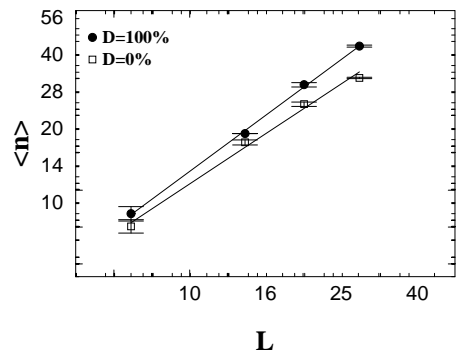}}
\end{center}
\vspace {-0.7cm}
\caption{Total number $\langle n\rangle$ of bonds cut during the whole breaking process as a function of size $L$ in log-log scale. For the strongly disordered case (filled circles) the slope is $1.14\pm 0.03$ and for the weakly disordered case (square) it is $1.02\pm 0.04$ }
\label{fig7}
\end{figure}

Figure 8 shows the relation between the average length of the crack $\langle M\rangle$ that percolates the system, causing its failure, and the size system $L$. Our results suggest the following scale behavior between $\langle M\rangle$ and  $L$
\begin{equation}
\langle M\rangle \sim L^{D_f},
\end{equation}
where $D_f$ is the fractal dimension. For cooper wires the value of the fractal dimension determined from least-squares fits is $D_f=1.03\pm0.03$  and for the steel wool wires  $D_f=1.05\pm0.04$. Taking into account the error bars we can conjecture that the exponent $D_f$ does not depend on material type. We have also verified that these results do not depend on the degree of disorder. Arcangelis and Herrmann \cite{arc1} also verified that for all the distributions considered, the length of the spanning crack that causes the failure scales with a fractal dimension of $D_f\sim1.1\pm0.1$.   

\begin{figure}[hbt]
\begin{center}
\resizebox{6cm}{!}{\includegraphics{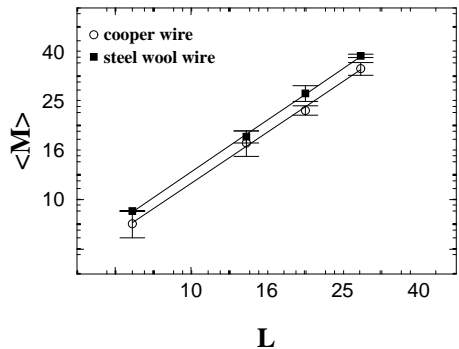}}
\end{center}
\vspace {-0.7cm}
\caption{Log-Log plot of the average length of the largest crack $\langle M\rangle$ as a function of the system size $L$. For the cooper wires (circle) the value of the slope is $1.03\pm 0.03$ and for the steel wood wires (full square) it is $1.05\pm 0.03$}. 
\label{fig8}
\end{figure}

It is important to point out that the process leading to the rupture of our network possesses characteristics of both the random fuse model \cite{arc1} and the dynamical thermal fuse model \cite{van,sorn}. In our experiments a change in the current distribution in the lattice causes a time evolution of the temperature of the fuses (similar to the dynamical thermal fuse model) but, due to the change of the conductivity of the wires, the current distribution changes again, until equilibrium is reached. The major difference between our experiment and the dynamical thermal fuse model is that in the latter the current is fixed at a certain value and the heating of the lattice due to generalized Joule heat leads, eventually, to its complete rupture. In our case the voltage is increased from zero and the time interval between successive voltage steps (few seconds) is short compared with the time that would take for the lattice to proceed to complete rupture, if left at that value of voltage. Then, only one or at most a few fuses burn in the time interval between consecutive voltage steps. This, in turn, makes our experiment similar to the simulations of the random fuse model, where only one fuse burns each time. Only when the voltage is very large is that a single step in its increase leads to a cascade of fuses burning, breaking the lattice completely. Preliminary results show that if the current is fixed at a value large enough, the rupture of our network proceeds in a manner similar to the one found by Lamaign\`{e}re {\it et al.} \cite{lamag}.

In conclusion, the results obtained by us confirm experimentally the predictions of the largely used random fuse model of Arcangelis and Herrmann \cite{arc1}. It was found that only for a strongly disordered system the current I, voltage V and the lattice size L obey the scaling law $I=L^\alpha(VL^{-\beta})$. Also, the results indicate the existence of two failure regimes, which depend on the degree of disorder: a catastrophic regime for $D<60\%$ and a disorder controlled regime for $D>60\%$. Both the total number of broken bonds and the average length of the crack have a power law dependence on the lattice size.

We thank A. J. M. Neves and S. O. Ferreira for helpful criticism of the manuscript. This work was partially supported by CNPq and FAPEMIG (Brazilian agencies).   
 
%

\end{document}